\documentclass[aps,prb,twocolumn,amsmath,showpacs,showkeys,superscriptaddress,floatfix]{revtex4}

\usepackage{amsmath}
\usepackage{epsfig}
\usepackage{graphicx}
\usepackage{graphics}
\usepackage{url}
\usepackage[english]{babel}
\usepackage{dcolumn}
\usepackage{color}
\begin{document}

\title{Optical signals of spin switching using the optical Stark effect in a Mn doped quantum dot
}

\author{D.~E.~Reiter}
\affiliation{Institut f\"ur Festk\"orpertheorie, Universit\"at M\"unster, Wilhelm-Klemm-Str.~10, 48149 M\"unster,
Germany}

\author{V.~M.~Axt}
\affiliation{Theoretische Physik III, Universit\"at Bayreuth, 95440 Bayreuth,
Germany}

\author{T.~Kuhn}
\affiliation{Institut f\"ur Festk\"orpertheorie, Universit\"at M\"unster, Wilhelm-Klemm-Str.~10, 48149 M\"unster,
Germany}


\date{\today}

\begin{abstract}
The optically induced spin dynamics of a single Mn atom embedded into a single semiconductor quantum dot can be strongly influenced by using the optical Stark effect. The exchange interaction gives rise to simultaneous spin flips between the quantum dot electron and Mn. In the time domain these flips correspond to exchange induced Rabi oscillations, which are typically off-resonant. By applying a detuned laser pulse, the states involved in the flipping can be brought into resonance by means of the optical Stark effect increasing the amplitude of the Rabi oscillations to one. In this paper we study theoretically how this spin dynamics can be monitored in time-resolved spectroscopy. In the spectrum the exchange interaction leads to a splitting of the exciton line into six lines, each corresponding to one of the six Mn spin states. The dynamical behavior of the Mn spin is reflected by the strength of the individual lines as a function of time. When an off-resonant optical pulse is applied the spectral positions of the lines shift, but still the flipping dynamics is visible.
\end{abstract}

\maketitle   

\section{Introduction}
When a single Mn atom is doped into a single semiconductor quantum dot (QD), possibilities for fascinating spin dynamics emerge. Due to the exchange interaction between the Mn spin and the electron and hole spin of the QD exciton, the energy levels of the states shift according to the Mn spin, leading in the case of a II-VI QD to the observed six line spectrum reflecting the six spin states of the Mn atom.\cite{besombes2004pro,fernandez-rossier2006sin, trojnar2011qua} Moreover simultaneous spin flips of the Mn spin and the electron (or hole) spin occur, which manifest themselves as anticrossings in the spectrum.\cite{besombes2004pro,goryca2010bri} Incoherent relaxation dynamics via coupling to phonons can be used to prepare the Mn spin in one of its extremal states, \cite{goryca2009opt1,legall2009opt, cywinski2010opt, cao2011spi} and also a coherent manipulation of the Mn spin via optical control of the QD excitons has been proposed.\cite{reiter2009all} We have shown that to flip the Mn spin by one either a series of $2\pi$ pulses or a strong magnetic field is needed. \cite{reiter2009all,reiter2009spi} With this we developed a protocol to switch the Mn spin from a given initial state into all other spin states. Using time-resolved spectroscopy the switching of the Mn spin can be monitored by the strength and spectral position of the absorption lines. \cite{axt2010ult_1}

To achieve flips of the Mn spin we recently proposed an alternative method,\cite{reiter2012spi} namely to use strong detuned laser pulses exploiting the optical Stark effect. \cite{legall2011opt} In this paper we discuss, whether the switching dynamics can be optically monitored in the presence of a strong, detuned pulse inducing the optical Stark effect.

\section{Theoretical background}
While a detailed description of our model can be found in previous works, e.g., in Ref.~\onlinecite{reiter2010opt}, here we summarize the main points. Our model of the QD accounts for the conduction band with electron spin $S^{e}=\frac{1}{2}; S^{e}_z=\pm\frac{1}{2}$, the heavy hole band with spin $S^{h}=\frac{3}{2}; S^{h}_z=\pm\frac{3}{2}$ and the light hole band with spin  $S^{h}=\frac{3}{2}; S^{h}_z=\pm\frac{1}{2}$. Heavy and light hole band are split by $40$~meV. A mixing between heavy and light holes has been neglected. From these bands 15 excitonic states are formed, namely the ground state $0$, the single exciton states $H\pm1$, $H\pm2$, $L\pm1$ and $L\pm0$, the combined biexciton states $HL\pm1$ and $HL\pm2$ and the biexciton states $HH0$ and $LL0$, where the number refers to the angular momentum of the exciton states and the letter to the valence band character of the involved states. Together with the six spin states of the Mn, $M_z= \pm\frac{5}{2}, \pm\frac{3}{2}, \pm\frac{1}{2}$, a basis of product states $|X,M_z\rangle$ is formed. The electron/hole spin and the Mn spin are coupled via the exchange interaction $j_{e/h}~\vec{S}^{e/h}\cdot \vec{M}$, which can be decomposed into an Ising type term $\sim S^{e/h}_z M_z$ and a flip-flop term $\sim \left(S^{e/h}_+ M_- + S^{e/h}_- M_+ \right)$.

The exchange coupling between electron and hole with the coupling matrix element $j_{eh}$ is modeled by an analogous interaction Hamiltonian. The coupling constants for the exchange couplings are $j_h=0.42$~meV, $j_e=-j_h/4$ and $j_{eh}=0.66$~meV.

When light excites the QD, the Mn spin state is not affected and only an angular momentum of $\pm1$ can be transferred to the exciton system. From the single exciton states only $H\pm1$ and $L\pm1$ are bright. We consider the usual light field coupling in dipole and rotating wave approximation. Using $\pi$ pulses with a Gaussian envelope and a width of $100$~fs, we excite or de-excite excitons in the system. The central frequency of the laser pulses is set resonant either to the bare heavy hole or light hole exciton transition energy ($E_{HH}$ or $E_{LH}$), i.e., not accounting for any coupling to the Mn. To make use of the optical Stark effect, the central frequency of the Stark pulse $\omega_{Stark}$ is increased by $5$~meV and the pulse envelope is taken to be rectangular with softened edges \cite{reiter2012spi}.

To visualize the spin dynamics in experimental accessible quantities, we model a time-resolved pump probe setup, where the QD is excited by a series of pump pulses consisting of the already introduced $\pi$ and Stark pulses and probed by a weaker pulse. The probe pulse has a width of $10$~fs and is delayed by $\tau$ with respect to the first $\pi$ pulse of the pump series. To calculate the probe spectrum we apply a widely used method, where a phase between pump and probe is introduced. The optical polarization is then expanded in a Fourier series with respect to this phase, where the Fourier coefficients correspond to different optical signals.\cite{haug1996qua,axt2004fem,axt2010ult_1}

\section{Dynamics in the presence of a Stark pulse}

\begin{figure}[t]%
\includegraphics[width=1.\columnwidth]{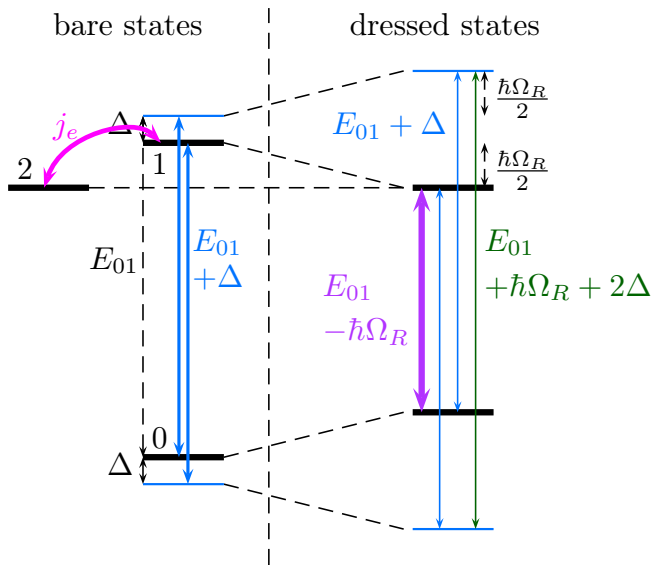}
\caption{(color online) %
	 Sketch of the three level system including interactions.}
\label{fig:sketch}
\end{figure}

\begin{figure}[t]%
\includegraphics[width=1.\columnwidth]{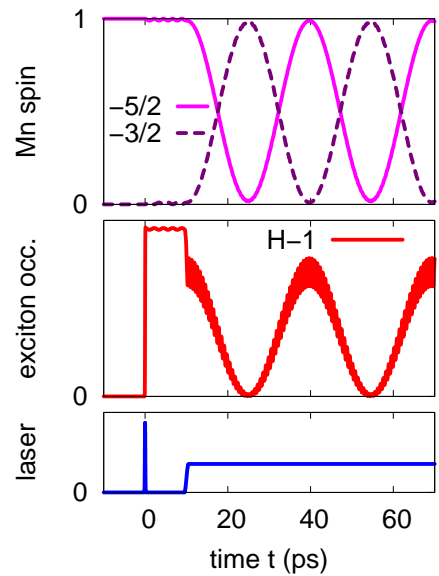}
\caption{(color online) %
	Occupation of the Mn spin states $M_z=-\frac{3}{2}$ and $M_z=-\frac{5}{2}$ (upper panel), occupation of the bright exciton $H-1$ (central panel), laser pulse (lower panel).}
\label{fig:1puls}
\end{figure}

In the initial state the QD is not excited and the Mn spin is taken to be extremal with $M_z=-\frac{5}{2}$. When a laser pulse excites the system, the Mn spin is unaffected and an exciton is generated. The exchange coupling can then induce spin flips between Mn spin and electron or hole spin depending on the type of the exciton. In the first step we focus on the excitation of the $H-1$ exciton. In this case the discussion can be reduced to the three level system sketched in left part of Fig.~\ref{fig:sketch}. The three levels are $|0\rangle = |0,-\frac{5}{2}\rangle$, $|1\rangle = |H-1,-\frac{5}{2}\rangle$ and $|2\rangle = |H-2,-\frac{3}{2}\rangle$, where the first two states are coupled via the light field, while the latter two states are coupled via the exchange interaction. The energy difference between state $|0\rangle$ and $|1\rangle$ is called $E_{01}$. The state $|1\rangle$ is shifted from the bare HH exciton energy by the Ising terms of the interactions by $2.2$~meV. Thus, the Stark pulse has an actual detuning of only $\Delta=2.8$~meV. The splitting between the states $|1\rangle$ and $|2\rangle$ is about $1.8$~meV. Systems with a periodic driving like the optical field can be interpreted in terms of steady states, where fast oscillating terms do not appear anymore. These states have a quasienergy, which is only well defined modulo $\hbar\omega_{Stark}$.\cite{sambe1973ste,harbich1981ele} In our case, this leads to a repetition of the bare states $|0\rangle$ and $|1\rangle$ at the energies $E_i+n\hbar\omega_{Stark}$. Two of these states additional to the three levels discussed above are marked in Fig.~\ref{fig:sketch} by thin blue lines, because they will be helpful to understand the effect of the Stark pulse.

At $t=0$ a $\pi$ pulse excites the system and at $t=10$~ps the Stark pulse sets in. The system dynamics can be followed in Fig.~\ref{fig:1puls}, where the occupation of the Mn spin states $M_z=-\frac{3}{2}$ and $M_z=-\frac{5}{2}$ (upper panel), the occupation of the bright exciton $H-1$ (central panel) and the laser pulse (lower panel) are shown. At $t=0$ we can see that the exciton state is excited. At the same time the occupations of the exciton as well as the occupations of the Mn spin states $M_z=-\frac{5}{2}$ and $-\frac{3}{2}$ start to oscillate weakly. These Rabi oscillations are driven by the exchange interaction and have a very small amplitude, because the energy splitting of the involved states is large. As soon as the Stark pulse sets in around $t=10$~ps the amplitude of the Rabi oscillation increases to almost $1$, because the energy of the bright exciton state is shifted into resonance with the dark state due to the optical Stark effect. In the occupation of $H-1$ an additional fast oscillation is observed, which is a light-induced, detuned Rabi oscillation driven by the Stark pulse. Its frequency is given by the splitting of the dressed states, which is $6.4$~meV, corresponding to an oscillation period of $0.65$~ps. Because the Mn spin is not affected by the light-induced Rabi oscillations, the fast oscillation is not seen in the occupations of the Mn spin states.

The calculations have been performed without taking into account dephasing. Therefore, let us briefly discuss the influence of pure dephasing on the switching process, which is typically the most important dephasing process in QDs at low temperatures and short times. For the excitation with the ultrashort $\pi$ pulse at $t=0$, pure dephasing should not play a role, because the excitation is much faster than the typcial time scale of the phonons \cite{krugel2005the}. After the excitation, the system is completely in the exciton state and no coherence between ground and exciton state, which would suffer from pure dephasing, is present. The situation is different during the action of the Stark pulse. Here, phonons can induce transitions between the dressed states, i.e. the eigenstates of the coupled system-light Hamiltonian, shown in the right part of Fig.~\ref{fig:sketch}. \cite{reiter2012pho, luker2012inf, wu2011pop, simon2011rob} However, since we start in the exciton state $|1\rangle$, we look at a transition from the lower dressed state to the upper dressed state, which would require the absorption of a phonon. The upper state is only weakly occupied, as seen by the weak amplitude of the Stark pulse-driven Rabi oscillations. At low temperatures phonon absorption processes are highly unlikely. Furthermore, the energy difference between the dressed states is $~6.4$~meV, which is much larger than typcial energies, where acoustic phonon processes are effective. \cite{gawarecki2012dep} Thus we conclude, that it is resonable to neglect influences of pure dephasing in our case.

\begin{figure}[t]%
\includegraphics[width=1.0\columnwidth]{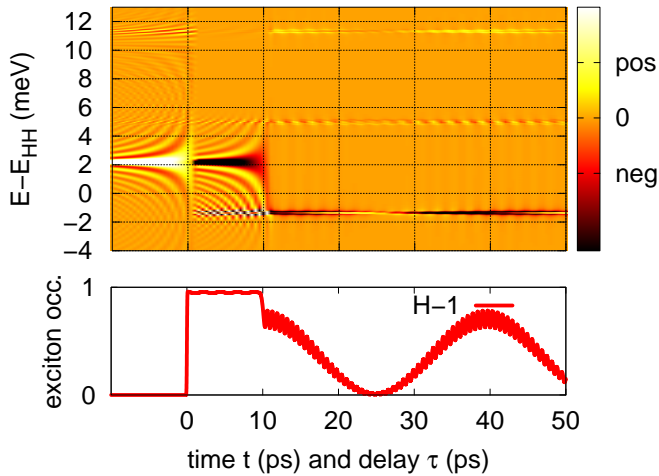}
\caption{(color online) %
	Contour plot of the probe spectrum as function of the delay $\tau$. The occupation of the exciton state $H-1$ is added as reference.}
\label{fig:PP}
\end{figure}

\begin{figure}[t]%
\includegraphics[width=1.0\columnwidth]{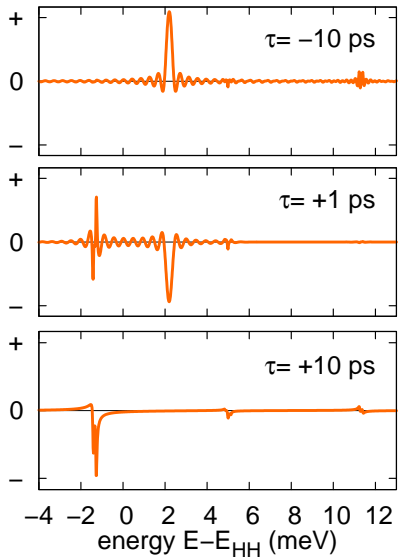}
\caption{(color online) %
	Probe spectra at $\tau=-10,1$ and $10$~ps.}
\label{fig:specs}
\end{figure}

The probe absorption spectrum corresponding to this dynamics is shown in Fig.~\ref{fig:PP}, where a contour plot of the spectrum as a function of the delay $\tau$ between pump and probe pulse is plotted. As a reference the occupation of the $H-1$ exciton is repeated in the lower panel of Fig.~\ref{fig:PP}. The spectra at selected delays $\tau=-10,1$ and $10$~ps are shown in Fig.~\ref{fig:specs}. Before the pump pulse the probe spectrum shows an absorption peak at $E-E_{HH}=2.2$~meV. This energetical position of the line corresponds to the Mn spin being in the state $M_z=-\frac{5}{2}$. Because for $\tau<0$ the probe pulse acts before pump pulse, the probe polarization is perturbed and modified by the action of the following pulse. This leads to the formation of the spectral oscillations overlayed on the absorption peak. The perturbation by the Stark pulse shifts the resonance energies of the system, which then influence the dynamics of the probe polarization. Therefore, already at negative delay times weak signatures appear at energies, which are present in the probe signal for times after the Stark pulse has been switched on. At $t=0$ the exciton is excited and correspondingly the peak changes its sign, indicating gain. The spectral position of the line is unchanged as the Mn spin has not changed. When the Stark pulse sets in at $t=10$~ps we see the line at $E-E_{HH}=2.2$~meV vanishing and a new feature at $E-E_{HH}=-1.4$~meV appears. The energy difference of the spectral line position is about $3.6$~meV and thus about twice as large as the energy difference between the states $|1\rangle$ and $|2\rangle$. In the spectrum the exchange induced Rabi oscillations between these states can be monitored by following the strength of the peak. The strength directly after the pulse is maximal. Then it decreases to zero, when half an oscillation has taken place at $t=25$~ps. Now the exciton is completely in the dark state and hence the system is transparent. When the system oscillates back the peak strength increases again to its maximal value at $t=40$~ps. If we look more closely at the spectrum at $\tau=10$~ps in Fig.~\ref{fig:specs}, we find that the Rabi oscillations are reflected by a double peak, because the coupling between the bright and dark state leads to an anticrossing. Additionally in the probe spectrum signatures with much weaker amplitude appear at $E-E_{HH}=11.4$~meV and at $E-E_{HH}=5.0$~meV.

The spectral position of the lines can be understood when we look at the sketch of the transitions in the right part of Fig.~\ref{fig:sketch}. Due to the interaction of the exciton system with the laser pulse, the states become dressed and their energies are modified by the instantaneous Rabi frequency $\Omega_R$ of the optical transition. \cite{allen1987opt} The Rabi frequency $\Omega_R$ should not be mistaken with the exchange induced Rabi frequency. This effect can be interpreted as a down shift of the energy of state $|1\rangle$ by $\frac{\Omega_R}{2}$ and an upshift of the energy of states $|0\rangle$ by the same amount. Correspondingly the transition energy lies at $E_{01}-\hbar\Omega_R$, which in our case is $E-E_{HH}=-1.4$~meV. The other peaks can be interpreted as transitions between quasienergy states as sketched as thin green and blue lines in Fig.~\ref{fig:sketch}, which have a splitting of $E_{01}+\hbar\Omega_R+2\Delta$, which calculates to $E-E_{HH}=11.4$~meV, and at $E_{01}+\Delta=E-E_{HH}=5$~meV in agreement with the peaks seen in Figs.~\ref{fig:PP} and \ref{fig:specs}.

\section{Mn spin flipping}

\begin{figure}[t]%
\includegraphics[width=1.00\columnwidth]{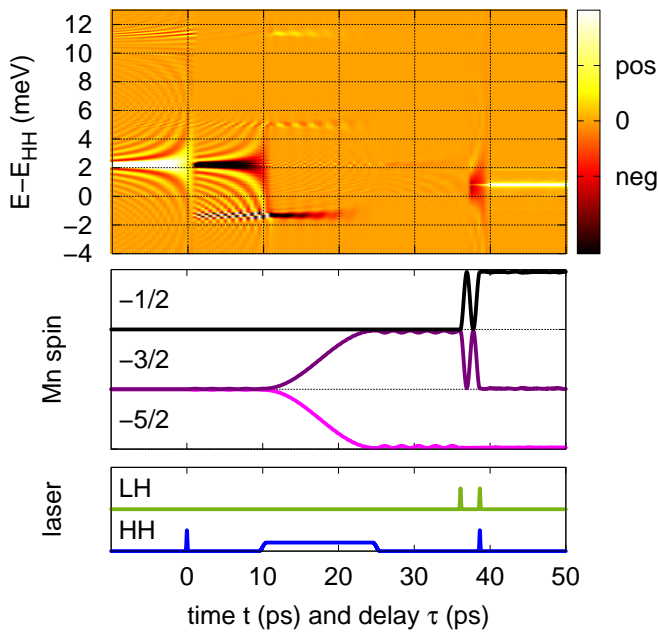}
\caption{(color online) %
	Contour plot of the absorption spectrum as function of delay $\tau$ (upper panel) under a pulse sequence (lower panel) to switch the Mn spin by two (central panel).}
\label{fig:PP2flip}
\end{figure}

Using the Stark effect it is also possible to switch the Mn spin by two, while excitons are only in the system during the switching time \cite{reiter2012spi} using our previously developed switching protocol. \cite{reiter2009all} For this protocol a combination of $\pi$ pulses resonant on the heavy hole (HH) and the light hole (LH) exciton transition are needed. In Fig.~\ref{fig:PP2flip} the pulse sequence using the Stark effect is shown (lower panel).\cite{reiter2012spi} The central panel in Fig.~\ref{fig:PP2flip} shows the occupation of the lowest three Mn spin states. We see that the Mn spin increases during the Stark pulse from $M_z=-\frac{5}{2}$ to $M_z=-\frac{3}{2}$. When $M_z=-\frac{3}{2}$ is reached the Stark pulse is turned off and a switch of the Mn spin by one is achieved. Then a $\pi$ pulse resonant on the LH exciton transition is applied and the combined biexciton $HL-1$ is created. Accommpanied by a LH spin flip the Mn spin increases to $M_z=-\frac{1}{2}$. Then by two $\pi$ pulses, one resonant on the HH transition and one resonant on the LH transition, the excitons are de-excited. Thus, an increase of the Mn spin by two from $M_z=-\frac{5}{2}$ to $M_z=-\frac{1}{2}$ has been achieved.

In the probe spectrum shown in the upper panel of Fig.~\ref{fig:PP2flip} we can follow this dynamics. For negative and short positive delays the spectrum is similar to the one shown in Fig.~\ref{fig:PP}. When the Mn spin has flipped by one and the Stark pulse is turned off, no signal is seen in the spectrum. Because the exciton is in a dark state, this state cannot contribute to the signal, but also blocks transitions to the bright exciton states. Thus the QD is transparent. When the combined biexciton is excited, a negative signal is seen. After the switching at $t=40$~ps, the exciton system is in the ground state again and we see a single absorption line at $E-E_{HH}=0.8$~meV. The spectral position of this line shows that the Mn spin is in the states $M_z=-\frac{1}{2}$, confirming the switch by two.

\section{Summary}
In this paper we have analyzed the optical signals corresponding to the optical manipulation of the Mn spin in a single QD using the optical Stark effect. The Stark effect is an efficient tool to switch between the Mn spin states.\cite{reiter2012spi} We have found that during the action of the Stark pulse, the energetical position of the lines shift according to the dressing of the states. The Rabi oscillations can be monitored by following the varying strength of the lines. A flip of the Mn spin by two can be seen by a change of the energetical position of the absorption line.


\end{document}